\documentclass[12pt]{article}
\usepackage{amsfonts}
\usepackage{amssymb}
\usepackage{graphicx}

\def\bC{{\mathbb C}}
\def\bR{{\mathbb R}}
\def\bP{{\mathbb P}}

\def\k{\kappa}
\def\hf{{1 \over 2}}

\def\G{\Gamma}
\def\Gp{\Gamma_{+}}
\def\Gm{\Gamma_{-}}

\def\tz{\tilde{Z}}

\newcommand{\be}{\begin{equation}}
\newcommand{\ee}{\end{equation}}
\newcommand{\bea}{\begin{eqnarray}}
\newcommand{\eea}{\end{eqnarray}}
\newcommand{\nn}{\nonumber\\}

\begin{document}

\begin{flushright}
\begin{tabular}{ll}
ITFA-2005-25\\
hep-th/0506230&
\\ [.3in]
\end{tabular}
\end{flushright}

\centerline{\Large \bf Knot invariants and Calabi-Yau crystals}
\centerline{}
\bigskip

\centerline{Nick Halmagyi$^{1}$, Annamaria Sinkovics$^{2}$ and Piotr Su\l kowski$^{2,3}$}

\bigskip
\centerline{
$^{1}$Department of Physics and Astronomy}
\centerline{University of Southern California}
\centerline{Los Angeles, California, 90089, USA}
\bigskip
\centerline{
$^{2}$Institute for Theoretical Physics, University of Amsterdam}
\centerline{
Valckenierstraat 65, 1018 XE Amsterdam, The Netherlands}

\bigskip
\centerline{
$^{3}$Institute for Theoretical Physics, Warsaw University}
\centerline{
ul. Ho\.{z}a 69, PL-00-681 Warsaw, Poland}

 \vskip .3in
{\it halmagyi@usc.edu}, {\it sinkovic@science.uva.nl}, {\it Piotr.Sulkowski@fuw.edu.pl}
\vskip .5in

\smallskip
 \vskip .3in \centerline{\bf Abstract}
 \smallskip

{We show that Calabi-Yau crystals generate certain Chern-Simons knot
invariants, with Lagrangian brane insertions generating the unknot and
Hopf link invariants. Further, we make the connection of the crystal brane amplitudes
to the topological vertex formulation explicit and show that the crystal naturally resums the 
corresponding topological vertex amplitudes. We also discuss the conifold and double
 wall crystal model in this context.  The results suggest that the free energy associated to
 the crystal brane amplitudes can be simply expressed as a target space Gopakumar-Vafa 
 expansion.}

\newpage	
 \section{{\Large Introduction}}

Recently, a quantum foam picture of topological string theory has been 
discovered \cite{ok-re-va, foam}. According to this duality, A-model 
topological string amplitudes on ${\mathbb C}^3$ and on more general toric 
Calabi-Yau manifolds can be computed by a statistical model of a melting 
crystal. The crystal is a physical picture of the A-model target space Kahler 
gravity, and as a quantum foam description, it captures the geometry up to 
very short distances.  The mathematical side of the correspondence is
the Donaldson-Thomas theory reformulation of Gromov-Witten invariants 
\cite{pand}.  

In particular the partition function of the melting crystal computes 
closed string A-model amplitudes. This was explicitly verified for ${\bC}^3$ 
\cite{ok-re-va}, and for more general non-compact geometries \cite{foam}.
This duality can be further extended by introducing non-compact brane probes
in the geometry. Such brane probes were found to correspond to defects in the 
$\bC^3$ crystal \cite{nat}.

It is also interesting to study brane probes in the crystal model of conifold
geometry. There are two ways to build a crystal for the conifold: in the
first way \cite{foam} one glues together two pieces of ${\mathbb C}^3$ 
geometries, in the second method of \cite{okuda} one makes use of the slicing
suggested by the open string description. It is the latter construction we
use in this paper. Here the crystal is like the ${\mathbb C}^3$ model, but 
ending in a wall in one direction. 

The conifold crystal is particularly interesting because it is a 
clear example of open-closed duality.  As is well-known, the closed 
topological A-model on the resolved conifold is dual to Chern-Simons theory 
on $S^3$ \cite{gv}. Further, the natural observables of  Chern-Simons theory 
are Wilson loop operators, related to knot and link invariants in the 
3-manifold $S^3$.  As is described in \cite{ov}, to each knot intersecting 
the $S^3$ we can associate a Lagrangian cycle, over which probe branes can be 
wrapped. Adding a Wilson loop observable along a knot in the Chern-Simons side 
thus corresponds to inserting non-compact Lagrangian brane probes in the 
closed string geometry.

It is then an interesting question how various Chern-Simons knot and
link invariants are encoded in the crystal model of conifold. The crystal
model is a simple statistical model of an infinite crystal with a wall in
one direction. Non-compact branes correspond to fermionic operators in the
transfer matrix formulation of the crystal, in agreement with the general
picture that non-compact D-branes in the topological B-model are fermions.
Operations in the crystal, such as the computation of amplitudes of non-compact 
branes are therefore easy. These amplitudes are then natural generating 
functions of certain knot invariants. 

In this paper we investigate the crystal picture of non-compact brane insertions
as generating knot expansions. Earlier work in this direction includes 
\cite{okuda}, where it was computed that insertion of a single 
brane corresponds to the unknot invariant in Chern-Simons theory, and related
observations about the connection of the crystal and topological vertex
formalism in \cite{nat}. 

In section \ref{crystal} we analyze the ${\bC}^3$ non-compact brane amplitudes.  
As ${\mathbb C}^3$ is the limit of the conifold when 
the Kahler parameter is sent to infinity, these amplitudes are generating  
functions of the leading part for certain knot invariants. In particular,
inserting two branes, one on each leg of the crystal generates the
leading part of Hopf link invariants. The more general case of many brane
insertions generates the leading part of Hopf link invariants in arbitrary representations.
In fact, we find that the many brane amplitudes can be viewed alternatively as generating
Hopf link tensor product representations,  corresponding to Young diagrams with a single row.
This latter point of view relates the crystal expansion to the topological
vertex formulation of the brane amplitudes explicitly. 

In section \ref{sec-conifold} we consider the conifold model of the crystal.
In particular,  we discuss how to generate the full unknot invariant in the conifold,
and derive the Ooguri-Vafa generating function.  
In section \ref{sec-CY-2walls} we introduce a crystal with two walls and compute the partition function 
of a single brane insertion. 

These non-compact brane amplitudes can also be derived in the topological
vertex formulation.  We compute and compare the same brane amplitudes 
in the A-model vertex formulation, where they are naturally expressed as knot 
expansions. We verify the crystal and vertex results agree. While the crystal framework is 
schematically simple to use, the summation of vertex 
amplitudes in many cases is complicated. The crystal then gives a simple and
natural closed expression for the vertex results. The comparison of crystal amplitudes with A-model topological vertex results is discussed in section \ref{sec-A-model}.
In section \ref{sec-B-model} we compare one nontrivial crystal amplitude with B-model topological vertex, also finding agreement.

Finally, section \ref{summary} contains a summary and discussion., where we consider
the connection of the crystal brane amplitudes (open Donaldson-Thomas invariants), 
Chern-Simons invariants and Gopakumar-Vafa invariants. In particular we conjecture that free 
energy associated to the crystal amplitudes can be simply expressed as a Gopakumar-Vafa 
expansion.  Thus D-brane degeneracies are simply encoded in the crystal free energy.

\section{Knot invariants from the crystal} \label{crystal}

The Calabi-Yau crystal is defined by a statistical sum over three dimensional partitions \cite{ok-re-va}, 
where  partitions are weighted by $q^{\#boxes}$, and $q=e^{-g_s}$. 

We will first consider the geometry $\bC^3$. In this case the crystal is understood as filling the positive octant of $\mathbb{R}^3$, which is a toric base of $\bC^3$.
One way to imagine the 3d crystal is to build from diagonal slices of two dimensional partitions. 
To assemble to 3d partitions, the diagonal slices have to satisfy the interlacing condition \cite{ok-re-va}.
A simple way to compute the crystal partition function is the transfer 
matrix formalism of \cite{ok-re-va}. In this formalism we assign a fermionic Fock space to each 
two dimensional diagonal slice. To construct the crystal in operator language, we use bosonization 
of the chiral fermion $\psi(z)= : e^{\phi(z)}:$, and the creation/annihilation part of the bosonic vertex operator, $\Gamma_{\pm}(z)$.  In this way the crystal partition function is built as \cite{ok-re-va}
$$
Z(q) = \sum_{3d~partitions} q^{\#boxes} = \langle 0| \prod_{m=1}^{\infty} \Gamma_{+}(q^{m-1/2})
\prod_{n=1}^{\infty} \Gamma_{-} (q^{-n+\hf}) \,|0 \rangle  \label{braneZ}
$$ 
From the commutation relations
$$
\G_{+}(z) \G_{-}(z')  = ( 1- z/z')^{-1}  \Gm(z') \Gp(z)
$$ 
it is straightforward to see that the partition function is the McMahon function $M(q)$
$$
Z(q) = M(q) = \prod_{n=1}^{\infty}\, (1 - q^n)^{-n}.
$$

In the B-model picture non-compact Lagrangian probe branes can be thought of as fermions
inserted in the geometry. These fermions are virtually free chiral fermions except they transform
between different patches with Fourier transformation \cite{ih}. The corresponding crystal description
of probe branes are fermionic operators \cite{nat} \footnote{For the crystal we use
p=0 framing.}
$$
\Psi_{D}(z) = \Gm^{-1}(z) \Gp(z).
$$
Similarly, anti-branes are represented by
$$
\Psi_{\bar{D}}(z)=\Gamma_{-}(z)\Gamma^{-1}_{+}(z). \label{anti-brane}
$$

\begin{figure}
\begin{center}
\includegraphics[width=0.45\columnwidth]{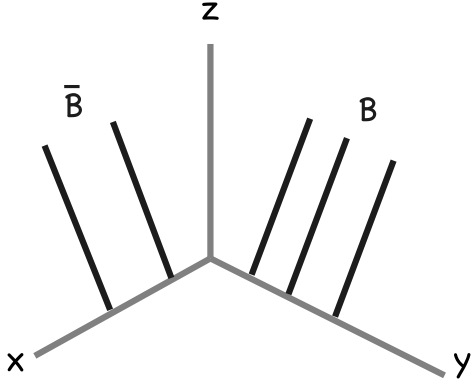}
\end{center}
\caption{\label{branesc3} \emph{Three Lagrangian branes inserted on the positive slice and two 
Lagrangian antibranes inserted in the negative slice of the toric geometry
of the ${\bC}^3$ crystal.}}
\end{figure}

A Lagrangian probe brane with geometry $S^1\times \bR^2$ (with crystal axis $(x,y,z)$)
$$
y = x + u = z + u, \qquad u= g_s (N+ 1/2) > 0
$$
ending on the $y$ axis at distance $u$ is described inserting a fermionic operator
$\Psi_{D,y}(e^{-u})$ at the slice $t=N+1$. Similarly, a brane at distance $v$ on the $x$-axis is
described by  $\Psi_{D, x}(e^{v})$ at the negative side of the diagonal $t=-(N+1)$ (Fig.~\ref{branesc3}).

Inserting $m$ Lagrangian branes at distances $g_s (N_i + \hf)$ ($i=1 \ldots m$)  on the $y$ axis, and $n$ Lagrangian anti-branes\footnote{We could of course have inserted branes, which would
cause a change of framing difference in the end result.} on the x-axis at distances $g_s(M_j + \hf)$,  ($j=1 \ldots n$),
as first derived in \cite{nat}, gives
$$
Z(a_1, \ldots a_n; b_1 \dots b_m; q) 
= \left\langle \, \Psi_{{\bar D}, x} (b_1) \ldots \Psi_{{\bar D}, x} (b_n)
\Psi_{D,y} (a_1) \ldots \Psi_{D,y} (a_m)\,  \right\rangle
$$
\be
= M(q) \left(  \prod_{i=1}^m \prod_{j=1}^n {L(a_i, q)   L(b_j, q) \over  (1 - a_i b_j)} \right) 
\left(\prod_{i>j} (1- a_i /  a_j) 
 (1- {b_i /  b_j}) \right). \label{nmbranes}
\ee
Here $L(a_i,q)$ for $a_i=q^{N_i+\hf}$ denotes the quantum dilogarithm
\be
L(a_i,q) = \prod_{i=1}^{\infty} ( 1 - q^{n + N_i}) = \sum_n a_i^n h_n(q^{\rho}), \label{def-dilog}
\ee
which can also be expressed in terms of the complete symmetric polynomials (defined in Appendix A).
The quantum dilogarithm is the brane wavefunction, as also can be seen from direct disk amplitude
computation \cite{diskm}, as well as from the insertion of a fermionic operator to the B-model geometry corresponding to the limit shape of the crystal \cite{ih}. The additional factors of type $(1-a_i /a_j)$
and $(1- a_i b_j)$ correspond to stretched strings between branes. We will explicitly see
later how these arise in the A-model topological vertex picture. 

In the following we will re-interpret this expression as a generating function of certain knot invariants in arbitrary representations.
As $\bC^3$ can be thought of as a limit of the conifold when its Kahler parameter $t =g_s N\rightarrow \infty$, by geometric transition we expect to see the leading part of knot invariants. 
We are then probing the invariants of 
$U(\infty)$ Chern-Simons theory.  More precisely, from 
the geometric picture of the Lagrangian branes with topology $S^1 \times \bR^2$ we expect to
find unknot and Hopf link invariants.
As the crystal result is written entirely
in terms of dilogarithms and simple prefactors from the stretched strings, it is not immediately
obvious that these would provide the generating functions for more complicated
link invariants, for example for Hopf link invariants in tensor product representations. 
It will turn out that the simplicity of crystal results is partly due to a particularly
natural framing choice.

\subsection{Single unknot}

Consider first a single brane on the $y$ axis. In this case, we have
$$
Z(a, q) = M(q) L(a, q).
$$
Normalized by $M(q)$, it is indeed the leading part of the generating function 
for unknot invariants as computed in Chern-Simons theory after the geometric 
transition \cite{ov}. It is also explicitly seen as a generating function
by expanding
\bea
L(a, q) &=& e^{\sum_{n=1}^{\infty} {a^n \over n [n]}} = 1 + {a \over 
(q^{\hf} - q^{-\hf}) } + a^2 {q^2 \over (q^2-1) (q-1)} + \ldots \nn 
&=& \sum_{R-one\ row} W_{R\bullet} a^{|R|}
\eea
Here the notation is $[n] = q^{n/2} - q^{-n/2}$, and the sum is rewritten 
as a sum over representations $R$. $W_{R\bullet}$ are the unknot invariants in zero
framing, and the summation runs over one row representations only, so that
$1=\square$, $2=\square\!\square$, etc. So a single brane inserted in the 
$C^3$ crystal computes the generating function of the leading part of unknot
invariants, for one row representations. 

\subsection{Hopf link}

Inserting an antibrane on $x$-axis and a brane on $y$-axis gives the generating
function of Hopf link invariants for single row representations.
Expanding the normalized part of the partition function
\be
{\tilde Z}(a, b, q) = {Z(a ,b ,q) \over M(q)} = 
{L(a,q) L(b,q) \over (1- a b)} \label{c3-antiDx-Dy}
\ee
gives
\be
\tz(a, b, q) =  \sum_{R, P\,-\,one\ row}  q^{{\k_{R} + \k_{P} \over 2}} W_{R^t P^t} a^{|R|} b^{|P|} \label{c3-antiDx-Dy-HL}
\ee
where $\kappa_R = |R| + \sum_{i} R_i (R_i -2i)$, for a general representation.
In the summation we only have one row representations. 
 We will later prove this expansion
by comparing with the topological vertex, and the q-dependent prefactors will be seen
as vertex framing factors $(-1, 0)$ : i.e. a brane at framing $-1$ and an antibrane at framing $0$. 
Alternatively, when expressed in terms of $q^{-1}$ this expansion gives Hopf Link coefficients with knot framing $(-1,-1)$.

\subsection{Hopf link with many rows} 

In the general case, for n branes on the y-axis and m antibranes on the x-axis 
the normalized partition function (\ref{nmbranes}) generates the leading part of Hopf link 
coefficients with $(n, m)$ rows

\bea
&& \tz(a_1, \ldots a_n; b_1, \ldots b_n;q) = \nn 
 && \sum_{R_1, \ldots R_n} \sum_{P_1, \ldots P_m}  
q^{{\k_{R} + \k_{P} \over 2}} W_{P^t R^t} a_1^{|R_1|} \ldots a_n^{|R_n|} 
b_1^{|P_1|} \ldots b_m^{|P_m|}  \label{claim-rhs}
 \eea
 where $R=(R_n, \ldots, R_1)$ and $P=(P_m,\ldots P_1)$ are $n$ and $m$ row representations
 respectively. 
The last summation also contains ``improper" Young-
diagrams.   For a proper Young diagram $(R_n, \ldots, R_1)$, we must have 
$R_1 \leq R_2 \ldots  \leq  R_n$.  Our summation contains also a finite number of terms where this
condition is not satisfied, but these improper contributions can still be formally
written using the definitions of Schur functions and Casimir $\kappa_R$.  Appendix B contains a partial proof of this formula (done for a simplified case) as well as details on the improper contributions.

While here it appears that the crystal generates Hopf link invariants in arbitrary 
representations, when comparing with the topological vertex, we will find the same
crystal partition function with a number of branes inserted on each leg can be viewed as 
generating more complicated link invariants, corresponding to the tensor product 
representations of Hopf link in one-row representations. 
We will return to this point in section \ref{sec-A-model}, 
where the crystal partition function as a knot generating function will be re-examined.

\section{Conifold crystal} \label{sec-conifold}

\begin{figure}
\begin{center}
\includegraphics[width=0.45\columnwidth]{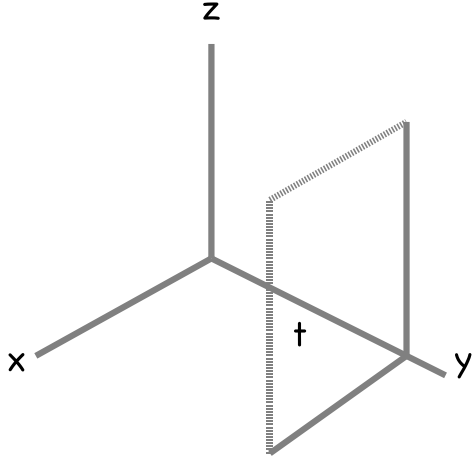}
\end{center}
\caption{\label{wall1} \emph{The toric geometry of the conifold crystal ending in a wall at the 
$y$ axes at distance corresponding to the Kahler parameter $t$.}}
\end{figure}

In the following we will examine how to obtain knot invariants from
the crystal model of resolved conifold $O(-1)\oplus O(-1) \rightarrow
{\bP}^1$. The crystal melting model describing topological A-model
on this geometry was obtained using the large $N$ dual Chern-Simons theory 
in \cite{okuda}. The geometry of the crystal reflects the toric diagram
of the resolved conifold, and it is obtained by inserting a wall in one
 direction. We will insert the wall at the positive slice $N$, constructing
the conifold geometry with Kahler parameter $t=g_s N$, which we often refer to as $Q=e^{-t}=q^N$
(Fig.~\ref{wall1}).

The partition function of this crystal model is thus obtained as
\begin{equation}
Z^{P^1}(q,N)=\langle 0|\prod_{m=1}^{\infty} 
\Gamma_{+}(q^{m-1/2})\,\prod_{n=1}^{N}
\Gamma_{-}(q^{-(n-1/2)})|0\rangle = 
M(q)\,e^{-\sum_{k}\frac{Q^{k}}{k[k]^{2}}} \label{P1-formula}
\end{equation}
in agreement with the topological vertex result (\ref{Z-P1}) \cite{vertex, marcos}.
Taking the Kahler parameter $t \rightarrow \infty$ gives back the 
partition function of the $\bC^3$ crystal.
Non-compact Lagrangian branes in the crystal are again defects described by
fermionic operators.

\subsection{Full unknot invariant}

Unknot invariants with many row representation can be generated by inserting 
a number of branes on the non-compact leg of the conifold crystal. 
This is analogous to the topological vertex picture as will be seen in
section \ref{sec-A-model}. Since now we have the full conifold geometry, we
will get the full unknot invariants, unlike in the $\bC^3$ geometry which
could only see the leading part of knot invariants (with $t\rightarrow 
\infty$).

Including $m$ antibranes at positions $a_i=q^{N_i+1/2}$, 
$i=1 \dots m$, at the non-compact leg 
their normalized partition function can be written as \footnote{Here 
normalization is with the conifold partition function and additional
$\xi(q)= \prod_{i=1}^{\infty} 1/(1-q^i)$ factors which has to be dropped
in comparison with topological string amplitudes.}
\be
\tz^{P^1}_{D}(a_1,\ldots a_n) = \Big[
\prod_{i<j}^{m}(1-\frac{a_i}{a_j})\Big] \prod_{i=1}^{m} \frac{L(a_i,q)}{L ( a_i Q,q )}.
\label{p1-brane-noncompact}
\ee
Taking a single brane first at $a=q^{N_1+1/2}$ gives the full 
unknot generating function for single row representations 
by the rearrangement
\bea
\tz^{P^1}_D(a) &=&  \frac{L(a,q)}{L ( aQ,q )} = \sum_{n=0}^{\infty} a^n \Big(\sum_{i=0}^n h_i(q^{\rho}) h_{n-i}(Qq^{-\rho}) 
\Big) = \nn
&=&  \sum_{n=0}^{\infty} a^n h_n(q^{\rho},Qq^{-\rho}).\label{conifold-brane}
\eea
In the first equality the expression of dilogarithm in terms of symmetric
polynomials is used given (\ref{def-dilog}), 
in the second
equality (\ref{schur-sum1}) was used. The final coefficient
$h_n(q^{\rho},Qq^{-\rho})$ is precisely the full unknot invariant 
(quantum dimension) for one-row representation, $R=(n,0
\ldots 0)$.
Taking $m$ branes and expanding in their positions $(a_1, \ldots a_m)$
gives similarly unknot invariants with $m$-row representation. The proof of
this is completely analogous to the induction included in Appendix B.

We note that the full unknot invariants were extracted before in \cite{okuda},
following a different prescription based on branes inserted in the compact
leg of $\bP^1$. Our procedure is different and is motivated by the topological
vertex picture as will be discussed in more detail below.

\subsection{Ooguri-Vafa generating function} \label{oogurivafa}

Chern-Simons theory on $S^3$ is the large $N$-dual to closed topological string theory
on the resolved conifold. The duality can be seen as a geometric
transition \cite{gv} - wrapping a large number
of branes on the base $S^3$ of deformed conifold, in the large $N$ limit the geometry transits to
the resolved conifold without branes. 

The geometry can be probed by non-compact branes \cite{ov}.
Wrapping the probe branes on a Lagrangian cycle, intersecting the $S^3$ in
a given knot, the worldvolume theory on the probe branes will also be a
Chern-Simons theory. In addition, there are open string stretched between 
the probe branes and the original large number of branes wrapping the $S^3$
and making the geometric transition. Integrating out these degrees of freedom
gives an effective theory on the probes branes, which is Chern-Simons
theory with additional corrections - the Ooguri-Vafa generating function.
For a single unknot it is given as \cite{ov}
$$
Z_{{\rm OV}} = \exp{ \left[- \sum_{n=1}^{\infty} {(e^{{n t \over 2}} - 
e^{{-nt \over 2}})
\over n [n]}  a_{{\rm OV}}^{-n} \right] }
$$
where $[n]=q^{n/2} - q^{-n/2}$ as before, and $a_{OV}$ is the parameter of the
one-dimensional holonomy matrix, that is the integral of holonomy of the
Chern-Simons gauge field around the circle loop (corresponding to the unknot) 
intersecting the $S^3$. After analytic continuation\footnote{Upper index 
$a$ denotes analytic.}, 
\be
Z^{a}_{{\rm OV}} = 
\exp{ \left[ {(a_{{\rm OV}}^n + a_{{\rm OV}}^{-n}) \over n [n]} 
e^{- n t/2} \right] } 
\ee
the Ooguri-Vafa generating function agrees with the topological string 
amplitude of a probe brane inserted in the conifold geometry. The
topological amplitude can also be derived considering the relevant open
topological string amplitude from the M-theory point of view of 
\cite{gv1}. Alternatively, it can be computed in the topological
vertex formulation. We will consider the latter computation in section
\ref{sec-A-model}.

Here we show that inserting a brane in the compact leg of the conifold 
crystal reproduces the Ooguri-Vafa generating function. Inserting an
antibrane\footnote{An antibrane is chosen for convenience here. When matching the crystal to the 
topological vertex result, we will choose the convention $q_{vertex} =1/q_{crystal}$, which turns an 
antibrane in the crystal to a brane in the vertex.}  on the compact leg of the crystal at the positive slice at $a=q^{N_0 +1/2}$
we obtain
\bea
&& Z^{P^1}_{D,y}(q,N_0, N) = 
\xi(q)\,M(q)\,e^{-\sum_{n>0}\frac{q^{n(N+1)}}{n[n]^{2}}}\,
e^{\sum_{n>0}\frac{q^{n(N_{0}+1/2)}+q^{n(N-N_{0}+1/2)}}{n[n]}} \nn
&=& \xi(q)\,Z^{P^1}\,L(a,q)L(Q/a,q),
\label{p1-brane}
\eea
where $Z^{P^1}$ is given in (\ref{P1-formula}), and 
now the Kahler parameter gets shifted to $t=g_s(N+1)$ due to brane insertion, so that $Q=q^{N+1}$. This is indeed the Ooguri-Vafa generating function with the identifications
$$
a_{{\rm OV}} = q^{N_{\rm OV}+\hf} \quad N_{{\rm OV}} = N_0 - {N \over 2}
$$
i.e. the position of brane in the geometry is measured from the middle point 
of the compact leg. We note that the crystal provides a straightforward
way to compute this result.

Inserting more branes on the compact leg would correspond to inserting more
stacks of branes in the geometry. The generating function can be easily 
computed on the crystal side. On the other hand,  in the crystal geometry it is not clear how to incorporate increasing the number of
branes in a single stack (thus increasing the holonomy matrix of probe).

It is a natural question to ask if inserting a number of branes on each 
leg of the conifold crystal would provide complete Hopf link invariants
with many rows, similarly to the leading part of Hopf link invariants
obtained from $\bC^3$. For example, inserting a brane on the compact leg
at position $a$ and an antibrane
on the non-compact leg at position $b$ in the conifold crystal 
one gets
$$
Z^{P^1}_{D;\,\bar{D}}(q,N) = Z^{P^1}{L(a,q) L(b, q) L(Q/a, q) \over L(bQ, q) (1 - ab)},
\label{p1-Dy-antiDx}
$$ 
where again the Kahler parameter gets shifted to $t=g_s(N+1)$ due to brane insertion on compact leg.
Expanding in $a$ and $b$ does not naturally give many row Hopf link invariants.
The reason is seen better in the language of topological vertex, where Hopf
link invariants are associated to having two branes inserted, each
on a non-compact leg of the conifold geometry \cite{marcos}. In the conifold crystal model
a Hopf link would naturally arise from placing branes on the non-compact
$x$-axis and another on the non-compact $z$-axis. In the diagonal slicing
of the crystal we work in, the latter branes are not natural to insert.
Working out the operators for insertion of such branes, and generating full 
Hopf link invariants from the crystal is left for future work. 

\section{Calabi-Yau crystal with two walls} \label{sec-CY-2walls}

\begin{figure}
\begin{center}
\includegraphics[width=0.45\columnwidth]{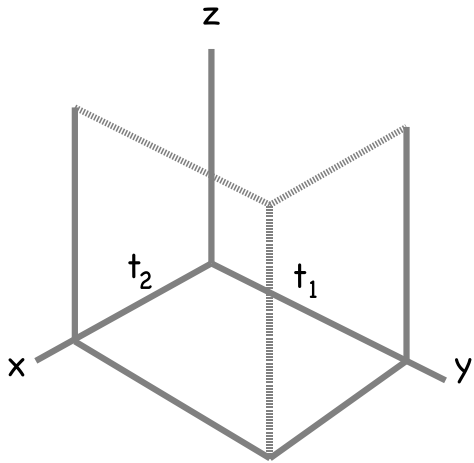}
\end{center}
\caption{\label{doublewall} \emph{The toric geometry of the crystal model of $\bP^1\times \bP^1$,
with two walls ending at the distances corresponding  corresponding to the Kahler parameters $t_1$
and $t_2$.}}
\end{figure}

The local conifold model for the crystal of \cite{okuda} can be easily 
generalized to represent the geometry with two neighbouring $\bP^1$.
This is naturally described by a crystal with two walls, on both the
positive and negative slice, at distance $t_1 =N_1 g_s$ and 
and $t_2=N_2 g_s$ respectively, which are the two Kahler parameters 
of the geometry (Fig.~\ref{doublewall}).

The partition function is computed as
\begin{eqnarray}
Z^{2\, walls}(q,N_1, N_2) & = & \langle 0|\prod_{n=1}^{N_1}\Gamma_{+}(q^{n-1/2})\,\prod_{m=1}^{N_2}\Gamma_{-}(q^{-(m-1/2)})|0\rangle = \nonumber \\
& = & \exp \sum_{k>0} \frac{(1-q^{kN_1})(1-q^{kN_2})}{k[k]^2}, \label{crystal-P1P1}
\end{eqnarray}
The factors in the exponent represent (apart from the unity giving McMahon function) worldsheets wrapping each of the spheres independently, and then both of them simultaneously.

Let us now insert a brane on the right compact leg at a position given by $a=q^{N_0+1/2}$, which
gives
\be
Z^{2\, walls}_{D,y} = Z^{2\, walls}\,\xi(q)\, L(Q_1/a, q )\,\frac{L(a,q)}{L(a Q_2,q)}, \label{crystal-D-P1P1}
\ee
where again there is a shift of the Kahler parameter corresponding to the leg the brane is put on.
It is also interesting to compare this result with a brane in the resolved conifold (\ref{p1-brane}). The essential difference is the dilogarithm 
in the denominator, which represents worldsheet wrapping a part of right 
$\mathbb{P}^1$ (of length $N_0$) and the whole left $\mathbb{P}^1$ 
(of length $N_2$).

It would be very interesting to generalize this construction by gluing 
together pieces of crystals to get an arbitrary toric geometry.

\section{Comparison with the topological vertex} \label{sec-A-model}

In this section we show that the amplitudes computed by crystal models with one or two walls, and multiple brane insertions, are indeed consistent with the
topological string results. We will perform the topological vertex calculations and 
appropriately match vertex and crystal moduli, and find a perfect dictionary 
between these two points of view. Most vertex calculations are performed in 
A-model language \cite{vertex}, but we also provide one nontrivial example of a B-model 
amplitude \cite{ih}.

Thus, let us focus on the A-model topological vertex. In this formulation the target space is a non-compact toric Calabi-Yau 3-fold, and topological amplitudes can be computed from a planar ``Feynman diagram" which encodes the geometry of the 3-fold. Each edge of such a diagram corresponds to a shrinking cycle of a toric fibration, and compact intervals represent local $\mathbb{P}^1$'s in Calabi-Yau geometry. The A-model vertex is a trivalent vertex for such a diagram and it encodes the structure of topological string in a single $\bC^3$ patch. The full toric 3-fold can be built from $\bC^3$ patches, and the amplitudes can be found by gluing vertexes according to the relevant gluing rules. The gluing process is implemented by a careful analysis of open strings ending on stacks of Lagrangian branes put on the axes. The vertex is most conveniently expressed in a representation basis as $C_{R_1R_2R_3}$, with each representation corresponding to a stack of branes on a single axis of $\bC^3$ patch. 

Apart from gluing string amplitudes, the vertex allows also to compute open string amplitudes in presence of a particular class of special Lagrangian branes of topology $\mathbb{C}\times \mathbb{S}^1$. The projection of these branes onto the plane of a toric diagram is a semi-infinite line with its endpoint attached to one edge of the diagram. For example, the partition function for inserting 3 non-compact branes on each leg of $\bC^3$
is computed as
$$
Z(V_1, V_2, V_3) =  \sum_{R_1, R_2, R_3}  C_{R_1, R_2, R_3}\, {\rm Tr}_{R_1} V_1\, {\rm Tr}_{R_2} V_2\, {\rm Tr}_{R_3} V_3\, , 
$$
where $V_i$ are sources (holonomy matrices) corresponding to inserted branes, 
and in general can be given by infinite matrices. This amplitude is written in the so-called canonical framing. In general, the vertex exhibits a framing ambiguity, which is a statement that one needs to specify one integer for each stack of branes to fully determine the amplitude. This is intimately connected with framing ambiguity in knot theory, and can be traced by a derivation of the vertex from Chern-Simons theory. All necessary details about computational framework for A-model, including framing ambiguity and other subtleties, are given in appendix C.

To match crystal and vertex results, a few important issues have to be taken into account. Firstly, the topological vertex is normalized in such a way that the McMahon function $M(q)$ of $\bC^3$ does not arise from calculations. Secondly, we need to choose some particular framing; usually this is $(-1)$ framing on one leg and the canonical one for other legs. Then, we have to take holonomy matrices $V_i$ to be one dimensional 
$$ 
V_i=a_i=q^{N_i+1/2} \label{brane-at-a}
$$ 
so in this sense the crystal can see only a fraction of what the full vertex computes. On the other hand, the crystal calculations are much simpler, so this is quite an advantage of using it.

If we have a single brane on one leg, then the above $a_i$ become simply moduli seen
in the crystal.
For more branes on one leg, we have to introduce parameters which give their positions, which must be combined with holonomy matrices
appropriately. We will see examples of this in what follows.

Finally, we perform  substitution 
\be
q\to \frac{1}{q}=q_{crystal} \label{inverse-q}
\ee
in vertex result to map crystal-branes to vertex-branes. In fact, in topological string such an operation exchanges branes to antibranes \cite{vertex}, and what we call branes and antibranes can be regarded just as a convention. Not performing the $q$ inversion would result in mapping crystal branes to vertex antibranes. We choose the former point of view. In fact, the $q$ inversion is important only for configurations with branes; 
the partition functions without any branes is invariant under $q\to 1/q$.

Let us note, that while some of the topological vertex amplitudes we consider here were already
written in the literature; it is not at all obvious that these amplitudes given by topological vertex rules in terms of sums over representations can be resummed into compact expressions, involving just dilogarithms and simple polynomials (as we have seen from crystal point of view). This fact was also
noticed in \cite{allcrystal}. Nonetheless, with the proper vertex framing chosen we rederive all these crystal results (which are in crystal canonical framing).

By construction, the topological vertex includes the correct worldsheet instantons which can appear in any toric construction, with or without Lagrangian probe branes. The contributions from specific instantons which stretch between probe branes can be read off from the form of the free energy. Specifically, the ${\rm Li}_1$ function in the factor

\be
(1-a b)={\rm exp} \left( {\rm Li}_1 (ab)\right),
\ee
appearing in all calculations involving more than one probe brane, shows that this is a contribution from annuli instantons and not of any higher genera instantons.


\subsection{Resolved conifold results}

We shall first test the conifold crystal results, the $\bC^3$ crystal results will
naturally follow then from taking the Kahler parameter to infinity. The 
resolved conifold partition function $Z^{P^1}$ (\ref{Z-P1}) has been computed before 
in several places \cite{foam, marcos}, and it is in agreement with the crystal result. 

Let us then compute brane configurations corresponding to those found in the crystal language. We start with a single brane on the external leg of the conifold in 
canonical framing
\begin{equation}
Z^{P^1}_{D-ext}(V) = \sum_{P,R} C_{\bullet R^{t} P} (-Q)^{|R|} C_{R\bullet \bullet}\, Tr_{P}V. \label{Z-brane-full}
\end{equation}
Using identities on Schur functions we get (see also \cite{marcos})
\begin{equation}
Z^{P^1}_{D-ext}(V) = Z^{P^1} \sum_P s_P(Qq^{-\rho},q^{\rho})\, Tr_{P}V. \label{Z-P1-brane-ext}
\end{equation}
Taking the matrix $V$ to be one 
dimensional $V=a=q^{N_0 + 1/2}$, and using $Tr_R(a)=s_R(a)$ and formula (\ref{schur-sum3}), we obtain
\be
Z^{P^1}_{D-ext} 
=  Z^{P^1} \sum_R s_R(Qq^{-\rho},q^{\rho})\, s_R(a)
=Z^{P^1} \frac{L(aQ,q_{crystal})}{L(a,q_{crystal})}.
\label{Z-D-L-v}
\ee
It is important to note, that the sum is in fact performed over representations
corresponding to diagrams with only one row (for a single number
$a$ and for any representation given by a diagram with more than one row
$s_{rep.\ with \ >1\ rows}(a)=0$). Taking into account the mapping (\ref{inverse-q}) we find perfect agreement 
with the normalized crystal result for antibranes (\ref{p1-brane-noncompact}).

A single brane can also be situated on the compact leg of the conifold at position $g_sD$
$$
Z^{P^1}_{D-int} = \sum_{R,Q^L,Q^R} C_{\bullet\bullet R\otimes Q^L} 
(-1)^s q^f e^{-L} C_{R^t\otimes
Q^R\bullet\bullet}\, Tr_{Q^L }V\, Tr_{Q^R}V^{-1}.  \label{comp-leg-brane-res}
$$
It is possible to perform resummation for $V=a=q^{N_0+1/2}$ and if $(-1)$ 
framing is chosen. 
If we follow the crystal convention and set the size of the compact leg to be $N+1$ (the shift  is responsible for brane insertion), and absorb the brane position into its modulus by defining 
\begin{equation}
N_0'=D+N_0, \qquad a'=q^{N'_0+\frac{1}{2}},
\end{equation}
we get after  substitution (\ref{inverse-q})
$$
Z^{P^1}_{D,y}=Z^{P^1}(N+1)\, L(a',q_{crystal})\,
L(Q/a',q_{crystal}), \label{brane-P1-comp}
$$
which is the same result as (\ref{p1-brane}).

It is also possible to insert several branes on the external or internal leg. For example, for $M$ branes on the compact leg in $(-1)$ framing we take $V_i=a_i=q^{N_i+1/2}$ and then get analogous factors as above. The Kahler parameter gets modified to $N+M$, and brane positions $D_i$ get absorbed similarly as above into $N_i'$ and modified moduli $a_i'$. We also have to take the stretched strings 
between the branes into account (see (\ref{stretched})).
All these factors combine to
\begin{equation}
Z^{P^1}_{M\,branes} =  {Z^{P^1}(N')} \Big[\prod_{i<j} (1-\frac{a_i'}{a_j'}) 
\Big] \Big[ \prod_{i=1}^{M} L(a_i',q_{crystal})\,
L(Q/a'_i,q_{crystal}) \Big], \label{M-branes-P1-comp}
\end{equation}
which is the same as the crystal result (4.42) in \cite{okuda}.

\subsubsection{Brane and antibrane on two legs}

\begin{figure}
\begin{center}
\includegraphics[width=0.5\columnwidth]{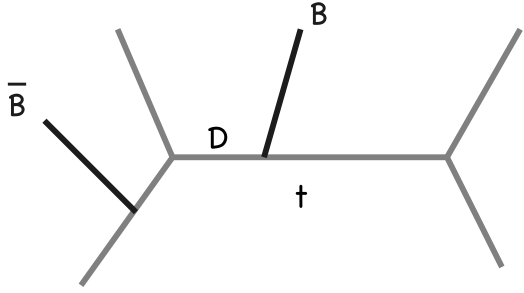}
\end{center}
\caption{\label{conifold1} \emph{The toric diagram of the conifold of Kahler parameter $t$, with a brane inserted at distance $D$ on the compact leg and an antibrane inserted on a non-compact leg.}}
\end{figure}

Let us put one brane on the compact leg of the resolved conifold at distance $D$ with holonomy matrix $V_1$, and the second brane on non-compact leg with holonomy $V_2$ (Fig.~\ref{conifold1}).

 We also take one-dimensional holonomy matrices $V_i=q^{N_i+1/2}$, and absorb the position on the compact leg into $V_1$
\begin{equation}
a=q^{D+N_1+1/2}=q^{N_1'+1/2}, \qquad b=q^{N_2+1/2}.
\end{equation}
The partition function in $(-1,0)$ framing is
\begin{eqnarray}
Z^{P^1}_{Dy;\bar{D}x} & = & \sum C_{R\otimes Q_L,P^t,\bullet} (-1)^{|P|} (-Q)^{|R|} C_{R^t\otimes Q_R\bullet\bullet} s_{Q_L}(a) s_{Q_R}(Q/a) s_{P}(b) \nonumber \\
& & \Big[(-1)^{|Q_L\otimes R|+|Q_R\otimes R^t|} q^{-\frac{\kappa_{Q_L\otimes R}+\kappa_{Q_R\otimes R^t}}{2}} \Big] = \nonumber  \\
& = & L\big(Q/a,q^{-1}\big) \sum s_{R^t}(-Qq^{\rho}) s_{Q_L}(-a) s_{P}(-b) \nonumber \\
& & c^{\alpha}_{R Q_L}  s_{\alpha^t/\eta}(q^{\rho}) s_{P^t/\eta}(q^{\rho}) q^{-\frac{\kappa_{R^t}}{2}}, \nonumber
\end{eqnarray}
where  the first dilog arises from $Q_R$ summation. Now summation over $P$ produces another dilog, and we can also sum over $Q_L$ and use (\ref{schur-invert2})
to get
$$
Z^{P^1}_{Dy;\bar{D}x}  =  L\big(Q/a,q^{-1}\big) L\big(b,q^{-1}\big) L(a,q^{-1}) \sum s_R(-Qq^{\rho}) s_{\eta /\alpha}(-a) s_{\eta}(-b) s_{R/ \alpha }(q^{\rho}),
$$
Performing the remaining sums over $R$, $\eta$ and finally $\alpha$ gives
the crystal result (\ref{p1-Dy-antiDx}) (after (\ref{inverse-q}) transformation) 
\begin{equation}
Z^{P^1}_{Dy;\bar{D}x} = \frac{Z^{P^1}}{1-ab} \frac{L\big(Q/a,q_{crystal}\big) L\big(b,q_{crystal}\big) L\big(a,q_{crystal}\big)}{L\big(bQ,q_{crystal}\big)}.
\end{equation}
Here we contrast the simplicity of crystal computation of the compact form 
final result to the extensive use of summation formulas and Schur identities 
in the above vertex computation.


\subsection{Double $\mathbb{P}^1$} \label{subsec-P1P1}

In this section we rederive two-wall crystal amplitudes from the topological vertex perspective. At first
 we compute the partition function. Denoting the sizes of the right and the left leg by $t_i$
(and $Q_i=e^{-t_i}=q^{N_i}$), respectively for $i=1,2$, the vertex rules and some rearrangements give
\begin{eqnarray}
Z^{P^1 P^1} & = & \sum_{P,R} C_{P^t\bullet\bullet} (-Q_2)^{|R|} C_{PR\bullet}
(-Q_1)^{|P|} C_{R^t\bullet\bullet} = \nonumber \\
& = & \sum_{\eta} \Big[\sum_{\mu} s_{\mu}(q^{\rho})s_{\eta}(Q_1 q^{-\rho})
s_{\mu}(Q_1 q^{-\rho}) \Big] \nonumber \\
& & \Big[\sum_{\nu} s_{\nu}(q^{\rho})s_{\eta}(Q_2 q^{-\rho}) s_{\nu}(Q_2 q^{-\rho})
\Big] = \nonumber \\
& = & \exp \sum_{k>0} \frac{-Q_1^{k}-Q_2^{k}+ (Q_1 Q_2)^k}{k[k]^2}, \label{P1P1}
\end{eqnarray}
This result is the same as the crystal expression (\ref{crystal-P1P1}), up to
McMahon function invisible for the vertex. Since this is a partition 
function without any brane insertions, it is also unaffected by $q$ inversion. 

\begin{figure}
\begin{center}
\includegraphics[width=0.36\columnwidth]{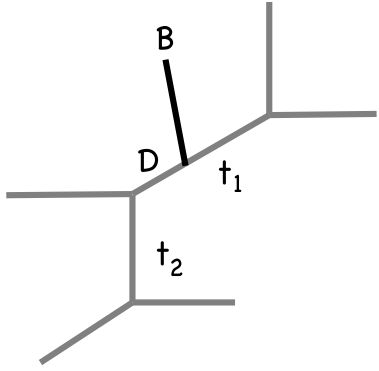}
\end{center}
\caption{\label{doublex} \emph{The toric diagram corresponding to a brane inserted in the geometry
$\bP^1\times\bP^1$ (with Kahler parameters $t_1$ and $t_2$) in the right compact leg at distance $D$
from the middle point.}}
\end{figure}

The vertex computation with a brane on the right compact leg of double 
$\mathbb{P}^1$, in (-1) framing also agrees with crystal result. Inserting 
this brane at position $D$ from the middle vertex (Fig.~\ref{doublex}), the topological vertex 
rules lead to the amplitude.

\begin{eqnarray}
Z^{P^1 P^1}_{D} & = & \sum C_{P^t\bullet\bullet} (-Q_2)^{|P|}  C_{R\otimes
Q_L,P,\bullet} (-Q_1)^{|R|}C_{R^t \otimes Q_R\bullet\bullet} \nonumber \\
& & q^{D|Q_L|} q^{(N_1-D)|Q_R|} \,Tr_{Q_L}V \,Tr_{Q_R}V^{-1} \nonumber \\
& &\Big[(-1)^{|Q_L\otimes R|+|Q_R\otimes R^t|}q^{-(\kappa_{Q_L\otimes R}+\kappa_{Q_R\otimes R^t})/2} \Big].
\end{eqnarray}
As before, we take one-dimensional $V=q^{N_0+1/2}$ and absorb the position into $V$ as
$$
a=q^{D+N_0+1/2}=q^{N_0'+1/2}.
$$
Performing sums over all representation in the appropriate order leads (after a little effort) to the result
\begin{equation}
Z^{P^1 P^1}_{D} = Z^{P^1 P^1} \frac{L\big(Q_1/a,q_{crystal} \big)
L\big(a,q_{crystal} \big)}{L\big(bQ_2,q_{crystal} \big)},
\end{equation}
which is the same as the crystal answer (\ref{crystal-D-P1P1}) after enlarging the
size of the right leg to $N_1+1$ due to the brane insertion. 


\subsection{$\mathbb{C}^3$ amplitudes}

The amplitude for several branes on one axis of $\mathbb{C}^3$ can be computed 
directly from the vertex rules, but since we already have the conifold result
it is easiest to take the  $N\to \infty$ limit in (\ref{M-branes-P1-comp}). This 
also gives result in $(-1)$ framing, and substituting (\ref{inverse-q}) we get
\begin{equation}
Z^{C^3}_{M\,branes}=\Big[\prod_{i<j} (1-\frac{a_i}{a_j})  \Big] \prod_{i=1}^{M}
L(a_i,q_{crystal}), \label{M-branes-C3-comp}
\end{equation}
which is the result for the $\mathbb{C}^3$ crystal, 
see (\ref{nmbranes}). For one brane it reduces to a single dilogarithm.

For a brane on one leg at position $a$ and antibrane on the other at
position $b$, and in framing $(-1,0)$, the vertex gives
$$
Z^{C^3\,vertex}_{D;\bar{D}}(a,b) = \frac{1}{1-ab} L(a,q_{crystal})  L(b,q_{crystal}).
$$
which reproduces the crystal answer (\ref{c3-antiDx-Dy}). In this case the vertex rules can be expressed in terms of Hopf link invariants (\ref{W-C})
$$
Z^{C^3\,vertex}_{D;\bar{D}}(a,b) = \sum_{P,R} W_{PR} (-1)^{|P|+|R|}q^{-\frac{\kappa_P+\kappa_R}{2}} s_P(a)s_R(b),
$$
so that inversing $q$ (\ref{inverse-q}) according to our conventions and using (\ref{W-PtRt}) proves that this is the same Hopf link generating function as in the crystal case (\ref{c3-antiDx-Dy-HL}). 

The calculation for two branes, one in
each leg, is similar and also gives the crystal result in $(-1,0)$ framing\footnote{
This is also an example of a situation, which can be resummed in canonical framing, with the final result $L(a,q) L(b,q) \frac{1-a\sqrt{q}+ab}{1-a\sqrt{q}}$.
This result does not agree with the crystal one (in canonical crystal framing),
thus a proper choice of framing is indeed crucial.}
$$
Z^{C^3\,vertex}_{D;\,D} = (1-ab)\frac{L(a,q_{crystal})}{L(b,q_{crystal})}.
$$

\begin{figure}
\begin{center}
\includegraphics[width=0.4\columnwidth]{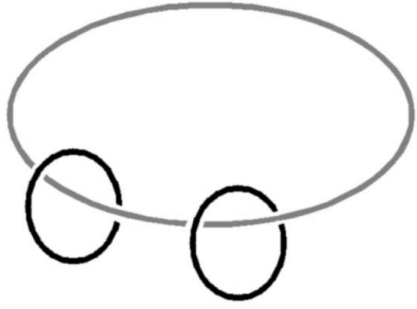}
\end{center}
\caption{\label{necklace2} \emph{The ``necklace" knot invariant generated by the insertion of
2+1 branes. The crystal only generates representations with a single row.}}
\end{figure}

The configuration with  two branes on one leg and antibrane on the other is slightly more complicated.
The stretched string factors between the two branes on the same leg, 
at positions $a_i=q^{M_i+1/2}$ (for $i=1,2$) give an $(1-a_1/a_2)$ factor. The 
full amplitude, with antibrane at $b=q^{N_1+1/2}$, and in $(-1,0)$ framing 
can be written as
\begin{eqnarray}
&& Z^{vertex}_{2Dy,\, \bar{D}x}  =  (1-\frac{a_1}{a_2}) \sum C_{P_1\otimes P_2, R^t,\bullet} s_{P^1}(a_1) s_{P_2}(a_2) s_{R}(b) (-1)^{|R|}  \times  \nonumber \\
&\times & \Big[(-1)^{|P_1 \otimes P_2|} q^{-\frac{\kappa_{P_1\otimes P_2}}{2}}  \Big] \nonumber \\
& = & (1-{a_1 \over a_2}){L(b,q^{-1})} \sum  c^{\alpha}_{P_1 P_2} s_{\alpha^t/ \eta}(q^{\rho}) s_{P_1}(-a_1) s_{P_2}(-a_2) s_{\eta}(-b) .  \label{necklace-resum}
\end{eqnarray}
After performing summations in several steps and substitution (\ref{inverse-q}) we recover the crystal result 
(\ref{nmbranes})
\begin{equation}
Z^{vertex}_{2Dy,\, \bar{D}x} =\frac{1-\frac{a_1}{a_2}}{(1-a_1 b)(1-a_2 b)}L(a_1,q_{crystal})L(a_2,q_{crystal})L(b,q_{crystal}) . \label{ncry}
\end{equation}
Thus  another way to look at the crystal result (\ref{ncry}) is provided by the first line in the expansion of (\ref{necklace-resum}), which due to (\ref{W-C}) can be written in terms of Hopf link invariants (with all components in knot $(-1)$-framing) as
\begin{eqnarray}
Z^{vertex}_{2Dy,\, \bar{D}x} & = & (1-\frac{a_1}{a_2}) \sum W_{P_1\otimes P_2, R,\bullet} s_{P^1}(a_1) s_{P_2}(a_2) s_{R}(b)  \nonumber \\
& & \Big[(-1)^{|P_1 \otimes P_2|+|R|} q^{-\frac{\kappa_{P_1\otimes P_2}+\kappa_{R}}{2}}  \Big]. \nonumber 
\end{eqnarray}
Taking out the stretched string factors $(1-a_1/a_2)$, the crystal result is seen as a generating function
 for  $2+1$ ``necklace" knot invariants. This knot is shown in
Figure \ref{necklace2}, arising from the tensor product representation of the Hopf link. Because of one-dimensional sources $V_i=a_i$, this is a generating function for representations with one row only. 

\subsubsection{Two legs of $\mathbb{C}^3$ - general situation}

Finally we consider $m$ branes on one leg at positions $a_i=q^{M_i+1/2}$, and $n$ antibranes on the next leg at $b_i = q^{N_i+1/2}$. 
As usual we take all branes in framing $(-1)$, which makes resummation doable. 
Using properties of tensor product, the part of the partition function without factors from strings stretching between branes on the same leg (\ref{stretched}) (which is denoted by ') can be written as
\begin{eqnarray}
Z'_{m,\bar{n}} & = & \sum C_{P_1\otimes \ldots \otimes P_m,\,R_1^t\otimes \ldots \otimes R_n^t,\bullet} (-1)^{\sum_i |R_i|} \Big[(-1)^{|\otimes_j P_j|} q^{-\frac{1}{2}\kappa_{\otimes_j P_j}} \Big]\cdot \nonumber \\
& & \cdot s_{P_1}(a_1)\ldots s_{P_m}(a_m)\cdot s_{R_1}(b_1)\ldots s_{R_n}(b_n) = \nonumber \\
& = & \sum s_{(P_1^t\otimes \ldots \otimes P_m^t)/\eta}(q^{\rho})  s_{P_1}(-a_1)\ldots s_{P_m}(-a_m)\cdot \nonumber \\
& & \cdot s_{(R_1^t\otimes \ldots \otimes R_m^t)/\eta} (q^{\rho})  s_{R_1}(-b_1)\ldots s_{R_n}(-b_n), \label{general-c3-vertex}
\end{eqnarray}
where it is understood that
$$
s_{(P_1^t\otimes \ldots \otimes P_m^t)/\eta} = \sum_{\alpha} c^{\alpha}_{P_1^t\ldots P_m^t} s_{\alpha/ \eta}.
$$
The antibrane part takes the form (here we write the partial result for
the $R$ summation only),
according to (\ref{multi-lit-rich}) 
$$
c^{\beta_{1}}_{R_1^t R_{2}^t} c^{\beta_{2}}_{\beta_{1} R_3^t}\ldots c^{\beta_{n-1}}_{\beta_{n-2} R_n^t} s_{\beta_{n-1}/\eta}(q^{\rho}) s_{R_1}(-b_1)\ldots s_{R_n}(-b_n) =
$$
$$
= s_{\beta_1^t/R_{1}}(-b_2) s_{\beta_{2}^t/\beta_{1}^t}(-b_3)\ldots  s_{\beta_{n-1}^t/\beta_{n-2}^t}(-b_n) s_{\beta_{n-1}/\eta}(q^{\rho}) s_{R_1}(-b_1) = 
$$
$$
= s_{\beta_{n-1}^t}(-b_1,\ldots,-b_n)s_{\beta_{n-1}/\eta}(q^{\rho}) = 
$$
\be
= L(b_1,q^{-1})\ldots L(b_n,q^{-1}) s_{\eta}(-b_1,\ldots,-b_n) \label{gen-c3-1}
\ee
In the same way, the brane part ($P$ summation separated) contributes
\be
L(a_1,q^{-1})\ldots L(a_m,q^{-1}) s_{\eta}(-a_1,\ldots,-a_m) \label{gen-c3-2}
\ee
The remaining summation over $\eta$ in (\ref{gen-c3-1}) and (\ref{gen-c3-2}) 
gives factors for strings stretched between all brane/antibrane pairs; also taking into account (\ref{stretched}) for each pair of branes (antibranes) on the same leg finally we get (after the q-inversion) 
\begin{eqnarray}
Z_{m,\bar{n}} & = & \Big[ \big(1-\frac{a_1}{a_2} \big)\ldots  \big(1-\frac{a_{m-1}}{a_m} \big) \Big]\Big[ \big(1-\frac{b_1}{b_2} \big)\ldots  \big(1-\frac{b_{n-1}}{b_n} \big)\Big] \nonumber \\
& & \frac{1}{1-a_1 b_1} \ldots  \frac{1}{1-a_{m} b_n} \prod_i L(a_i,q_{crystal})\ \prod_j L(b_j,q_{crystal}), \nonumber
\end{eqnarray}
and this is the same answer as we found from the crystal (\ref{nmbranes}). 

\begin{figure}
\begin{center}
\includegraphics[width=0.5\columnwidth]{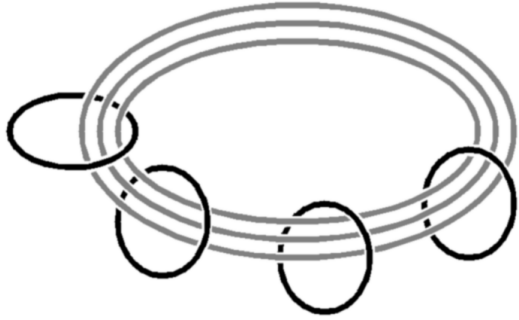}
\end{center}
\caption{\label{chain2} \emph{The ``necklace" knot invariant generated by the insertion of
4+3  branes in the crystal, corresponding to the Hopf link tensor product representation.}}
\end{figure}

This more general case also can be understood as a generating function of Hopf link invariants corresponding to tensor products of one-row representations, as the first line of (\ref{general-c3-vertex}) can be written using (\ref{W-C}) as
\begin{eqnarray}
Z'_{m,\bar{n}} & = & \sum W_{P_1\otimes \ldots \otimes P_m,\,R_1\otimes \ldots \otimes R_n}  \Big[(-1)^{|\otimes_j P_j|+|\otimes_k R_k|} q^{-\frac{1}{2}(\kappa_{\otimes_j P_j} + \kappa_{\otimes_k R_k})} \Big]\cdot \nonumber \\
& & \cdot s_{P_1}(a_1)\ldots s_{P_m}(a_m)\cdot s_{R_1}(b_1)\ldots s_{R_n}(b_n), \nonumber 
\end{eqnarray}
where factors from strings stretched between branes on the same leg (\ref{stretched}) are taken out. 
The corresponding knots are shown in Figure \ref{chain2}, for the case of four branes and three antibranes inserted in
the geometry.

Thus the crystal generating function can be
interpreted in two distinct ways, in the first way described in section \ref{crystal}
it is the generating function of Hopf link invariants for representations with several rows.
In the second way (as shown here from the topological vertex point of view) expanded without the stretched string
factors it generates necklace (or tensor product) knot
invariants with a single row in knot framing $(-1,-1)$.

\section{B-model example} \label{sec-B-model}

In the B-model \cite{ih} topological amplitudes are computed on the mirror 
Calabi-Yau geometries. The mirror geometry is described by the general equation 
$xy-F(u,v)=0$. To compute the B-model amplitudes we follow the formalism 
of \cite{ih} closely, where the B-model amplitudes are computed as
$$
\langle vac|(\,branes \,/\, antibranes\,)  |V\rangle, 
$$
where $\langle vac|$ is a vacuum state chosen in a way which ensures that overall fermion number is zero, and
$$
|V\rangle = \exp \sum_{k,l\geq 0} \Big(a_{kl}\psi_{-k-1/2}\psi^{*}_{-l-1/2} + \tilde{a}_{kl}\psi_{-k-1/2}\tilde{\psi}^{*}_{-l-1/2}    \Big) |0\rangle \label{V}
$$
is a state representing the Riemann surface $F(u,v)=0$ branes live on. This Riemann surface might have several asymptotic ends, with branes in each of them; we restrict ourselves putting branes in two of the patches. The quantities in these two patches are denoted without and with tilde respectively, and positions of branes are given by $e^{-u_i}=a_i$ and respectively $b_i$. In B-model picture branes are 
represented by fermions with standard mode expansions, thus in two patches we have
\bea
\psi(a) & = & \sum_k \psi_{k+1/2} \, e^{-(k+1)u_i} =  
\sum_k \psi_{k+1/2} a^{k+1}, \nn
\tilde{\psi}(b) & = & \sum_k \tilde{\psi}_{k+1/2} b^{k+1}. \nonumber
\eea
Only fermions from the same patch anticommute
$$
\{ \psi_{-k-1/2},\,\, \psi^*_{l+1/2} \} = \delta_{k,l},
$$
and the bare vacuum is annihilated by all positive modes
$$
\psi_{k+1/2}|0\rangle = \psi^{*}_{k+1/2}|0\rangle = \tilde{\psi}_{k+1/2}|0\rangle = \tilde{\psi}^{*}_{k+1/2}|0\rangle = 0 \qquad \textrm{for}\ k\geq 0.
$$
In the case of $\mathbb{C}^3$ the state $|V\rangle$ is determined (up to $q^{1/6}$ factors) by
\begin{eqnarray}
a_{kl} & = & (-1)^l s_{hook(k+1,l+1)}(q^{\rho}), \nn
\tilde{a}_{kl} & = & (-1)^l q^{-\frac{\kappa_{(l+1)}}{2}} (W_{k+1,l+1} - W_{k+1}W_{l+1}), \nonumber
\end{eqnarray}
where $hook(m+1,n+1)$ is a hook representation with the relevant number of boxes in its row and column, and $W_{k+1,l+1}$ is Hopf link invariant for two symmetric representations with relevant number of boxes. For symmetric representation, the value of Casimir is $\kappa_n = n^2 -n$.

\begin{figure}
\begin{center}
\includegraphics[width=0.5\columnwidth]{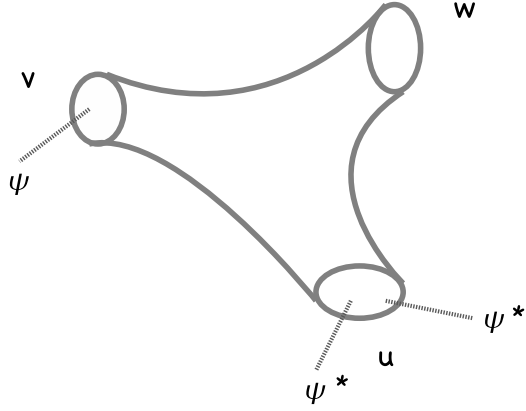}
\end{center}
\caption{\label{Bmodelpsi} \emph{Two antibrane and a brane inserted in two asymptotic patches of the mirror B-model geometry.}}
\end{figure}

Now we put two antibranes in one patch (these in framing $(-1)$) and a single brane in the other one (Fig.~\ref{Bmodelpsi}).\footnote{We could have started with two branes and
an antibrane, two antibranes and a brane is just slightly more convenient for
the B-model computations.} The vacuum should be chosen as $\langle vac| = \langle 0| \tilde{\psi}_{1/2}$, and in this case the only contribution comes from the third coefficient (with $1/2$ factor) in the exponent expansion of $|V\rangle$. Manipulations with fermion operators lead to \begin{eqnarray}
& & \langle vac|  \tilde{\psi}(b) \psi^*(a_1) \psi^*(a_2) |V^{(-1,0)}\rangle = \nonumber \\
& &  = \sum_{p,t,r \geq 0} \tilde{a}_{pt} \tilde{a}_{r0} \,b^{t+1}\big(a_1^{r+1}a_2^{p+1}- a_1^{p+1}a_2^{r+1} \big) (-1)^{-p-r}q^{-\frac{\kappa_{p+1}+\kappa_{r+1}}{2}}. \label{2Dbar-D-I} \nonumber
\end{eqnarray}
Performing the summation gives
$$
\langle vac|  \tilde{\psi}(b) \psi^*(a_1) \psi^*(a_2) |V^{(-1,0)}\rangle =  \frac{-a_1 a_2 b}{L(a_1,q)L(a_2,q)L(b,q)} \Big(\frac{1}{1-a_2 b} - \frac{1}{1-a_1 b}\Big)
$$
$$
= \frac{a_1 a_2^2 b^2}{L(a_1,q)L(a_2,q)L(b,q)} \frac{1-\frac{a_1}{a_2}}{(1-a_1 b)(1-a_2 b)}.
$$

We already know that inversing $q$ exchanges branes with antibranes. So if we started with two branes in the first patch and antibrane in the second, we would get dilogs in numerator. This agrees with the crystal result (\ref{nmbranes}), up to the irrelevant overall $a_1 a_2^2 b^2$ factor.


\section{Summary and discussion} \label{summary}

In this paper we investigated the appearance of knot invariants in
the construction of Calabi-Yau crystals. Inserting Lagrangian branes,
the $\bC^3$ crystal naturally generates the leading part of unknot and 
Hopf link invariants, with arbitrary number of rows. Comparison with the 
topological vertex gives an alternative view of the crystal generating
invariants for Hopf link for tensor product representations (Fig.~\ref{chain2})
with a single row.

\subsection{Connection to Gopakumar-Vafa invariants}

The connection to knot invariants is entirely expected from the topological 
vertex point of view, which is itself constructed from Chern-Simons knot invariants,
using open-closed duality. However, the crystal is interesting for 
it simplicity summing the vertex knot expansions in natural generating 
functions. These generating functions are always dilogarithms and simple
prefactors. We can phrase this as the statement that inserting branes in the crystal generates 
(open) Donaldson-Thomas invariants (related to open topological string amplitudes) 
and here we express these Donaldson-Thomas  invariants in terms of Chern-Simons invariants.
We stress the simplicity of the crystal computing these DT knot generating 
functions, as compared to other methods.

Open string topological amplitudes can also be derived from Gromov-Witten theory, by counting
holomorphic maps with boundaries in Lagrangian submanifolds. These amplitudes can
alternatively  be computed from the target space point of view using the M-theory perspective 
\cite{gv1}, where they contain information about counting of BPS states.   
Based on the geometric transition picture of the conifold, the Ooguri-Vafa generating function constructed from Chern-Simons invariants can be reformulated in  terms of BPS degeneracies
counting D2-branes ending on D4-branes \cite{ov} as\footnote{The OV conjecture is naturally formulated in the free energy rather than the partition function.}
\be
F_{OV} =  \sum_{n=1}^{\infty} \sum_{R, Q, s} {N_{R,Q,s} \over  n [n] }
e^{n ( - t_Q + s g_s )} Tr_{R} V^n  \label{gopakumarvafa}
\ee
where $N_{R,Q.s}$ are the BPS degeneracies labeled by representation, charge and spin content,
$[n]= q^{n/2} - q^{-n/2}$ as before;  $t_Q =\int_{Q} k$ is the area of the corresponding cycle, and $V$
is the holonomy matrix. For the case of the unknot in $S^3$ this precisely gives the
Ooguri-Vafa unknot  generating function 
$$
F_{OV} = 
\sum_{n=1}^{\infty} {Tr V^n + Tr V^{-n}  \over n [n]} 
e^{- n t/2}
$$ 
The connection between Chern-Simons and Gopakumar-Vafa invariants is elaborated in the series of works \cite{marcoscs}.

In the case of closed topological strings, Donaldson-Thomas invariants are new invariants 
which reformulate the Gromov-Witten theory physically in target space language.

Given that the Calabi-Yau crystal naturally computes the closed and open topological string amplitudes
in target space language, we certainly expect a natural relation to the partition function
and D-brane amplitudes computed by the Gopakumar-Vafa formulation.
In fact, this connection can be explicitly seen in our results already. The crystal brane amplitudes
are naturally given in terms of dilogarithms, and we can use the exponential expansion of the dilogarithm 
$$L(a,q) = e^{\sum_{n=1}^{\infty} {a^n \over n [n]}}$$ 
to extract the free energy. Recalling that our holonomy matrix is one dimensional, thus $Tr V$ is
related to $a_{OV}$, the free energies of the crystal brane amplitudes clearly are of the 
Gopakumar-Vafa form (\ref{gopakumarvafa}). This is explicitly checked by the computation of the
 Ooguri-Vafa generating
function inserting the brane in the conifold crystal in section \ref{oogurivafa}.

Since all of the brane amplitudes written similarly  in terms of dilogarithms, we conjecture the free 
energies obtained from the crystal
brane partition functions (extracted with the exponentiation formula) are
natural expansions in the Gopakumar-Vafa invariants. 
Thus according to this the crystal amplitudes also naturally compute D-brane degeneracies
(they can be simply read off from the expression of crystal free energy). This would all fit
in the point of view that the crystal Donaldson-Thomas theory is really a target space theory,
and as such it simply encodes the target space point of view of D-brane amplitudes.
It would be very interesting to explore the connection between the DT, Gopakumar-Vafa and
Chern-Simons invariants in more detail.

\subsection{Open questions}

There are several open questions related to our work.
In particular, we only investigated the simplest unknot and Hopf link invariants from the crystal
point of view. It would be very interesting to find how more complicated knots are generated from
the crystal. One way to realize this would be to investigate how skein relations are represented
in the crystal language. One can possibly also use the formalism of knot operators to find the crystal 
representation of more complicated knots, like for example torus knots. 
Since Lagrangian brane insertions only produce Hopf link and torus knot invariants, it would
be very interesting to understand if there are natural geometric objects (like combination of branes,
or new classes of branes) which would compute more complicated knots. 

Another question is how to represent the full topological open A-model 
amplitudes in the crystal. While in the topological vertex one inserts stacks of D-branes, in the
crystal we only used a single D-brane probe in each stack. That is  the holonomy matrix seen in
the crystal is one dimensional only, while in the topological vertex it can be arbitrarily large.
Finding the representation of holonomy matrix in the crystal would allow to compute the full structure
of A-model amplitudes, in particular that would give also multiple row tensor product Hopf link invariants.
Introducing the holonomy matrix may have to do with the generalized fermionic operators found
in the crystal in \cite{univ}.

Another important open  problem is how to extend the crystal amplitude computations for more complicated toric geometries. Clearly, one has to glue pieces of crystal geometries to study more complicated toric amplitudes than the double ${\mathbb P}^1$ geometry we studied in this paper.   
A gluing prescription for toric diagrams involving a partition function only is given in \cite{foam}.
It would be important to understand how to glue pieces of crystals with D-branes inserted.
One way to proceed in this direction is to take guidance from the topological vertex gluing
prescriptions, and the clear-cut relations we found between certain class of vertex and crystal
brane amplitudes in this paper. 

Finally, we note that the Chern-Simons model of the crystal may have a natural connection to
the Brownian motion picture of \cite{sebas}. It would be interesting to investigate this direction
further to find a string theory realization of this picture. 

\subsection*{Acknowledgments}
We are grateful to Robbert Dijkgraaf and Marcos Marino for enlightening discussions. We also
thank Nick Jones, Albrecht Klemm, Asad Naqvi, Takuya Okuda and Jacek Pawe\l czyk for valuable conversations. P.S. would like to especially thank Robbert Dijkgraaf for all the support, Amsterdam String Theory group for great hospitality and NWO Spinoza Grant for assistance. N.H. would like to thank the YITP at Stony Brook and the Research School for Theoretical Physics at ANU for hospitality. In addition N.H. is supported by a Fletcher Jones graduate fellowship from USC.
The work of A.S. is partially supported by the Stichting FOM. 

\appendix
\section*{Appendix A} \label{app-schur}

Many of our formulas make use of properties of symmetric functions. Here we summarize
the basic properties and some identities for symmetric functions: Schur
polynomials $s_{R}$, elementary $e_{R}$ and complete $h_{R}$ symmetric polynomials,
Newton polynomials $P_{R}$.

A symmetric polynomial $S$ depends on a partition $R$, and its argument
is a string of variables $x=(x_1,x_2,\ldots)$, what we denote by
\be
S_R(x)=S_R(x_1,x_2,\ldots).
\ee
By $q^{R+\rho}$ we understand a string such that $x_i=q^{R_i-i+1/2}$ for
$i=1,2,\ldots$, thus
$$
S_R(q^{R+\rho})=S_R(q^{R_1-1/2},q^{R_2-3/2},\ldots).
$$
In particular
\be
S_R(q^{\rho})=S_R(q^{-1/2},q^{-3/2},\ldots).
\ee

One can concatenate two strings of variables, $x=(x_1,x_2,\ldots)$ and
$y=(y_1,y_2,\ldots)$, and then use it as an argument of a symmetric polynomial,
which is denoted by
$$
S_{Q}(x,y)=S_{Q}(x_1,x_2,\ldots, y_1,y_2,\ldots).
$$

One of the simplest examples of symmetric functions are \emph{Newton polynomials}
\begin{equation}
P_R(x)=\prod_{n} P_{R_i}(x), \quad \textrm{where} \quad  P_n(x)=\sum_{i=1}x_{i}^{n}.
\label{newton}
\end{equation}

Let us next introduce \emph{elementary} $e_n(x)$ and \emph{complete symmetric
functions} $h_n(x)$, for $n=0,1,2,\ldots$, in terms of a generating functions
\begin{eqnarray}
E(t) & = & \sum_{n=0}^{\infty} e_n t^n = \prod_{i} (1+x_i t), \label{e-n} \\
H(t) & = & \sum_{n=0}^{\infty} h_n t^n = \prod_{i} \frac{1}{1-x_i t}, \label{h-n}
\end{eqnarray}
and $h_{-1}=e_{-1}=h_{-2}=e_{-2}=\ldots=0.$
Then, for a partition $R=(R_1,R_2,\ldots)$,
\begin{eqnarray}
e_R & = & e_{R_1}e_{R_2}\cdots \nonumber \\ \label{e-R}
h_R & = & h_{R_1}h_{R_2}\cdots. \label{h-R} \nonumber
\end{eqnarray}

For a partition $R$, the \emph{Schur function} is defined as
\begin{equation}
s_R(x)=\det(h_{R_i -i+j})=\det(e_{R^{t}_i -i+j}). \label{s-det}
\end{equation}

Let us introduce Littlewood-Richardson coefficients $c^{P}_{QR}$ as
\begin{equation}
s_{Q\otimes R}=s_Q s_R=\sum_P c^{P}_{QR}s_P, \label{lit-rich}
\end{equation}
which have properties
\begin{equation}
c^{P}_{QR} =  c^{P^t}_{Q^t R^t} = c^{P}_{RQ}, \qquad c^{P}_{R\bullet} = \delta^P_R,  \label{lit-rich-prop} \\
\end{equation}
\begin{equation}
c^{P}_{QR} = 0 \ \textrm{for}\ |P| \neq |Q|+|R|. \label{lit-rich-nonzero}
\end{equation}
It is also convenient to define multiple coefficient
\begin{equation}
c^{P}_{R_1\ldots R_n} = \sum_{\alpha_i} c^{\alpha_1}_{R_1 R_2} c^{\alpha_2}_{\alpha_1 R_3} c^{\alpha_3}_{\alpha_2 R_4} \cdots c^{P}_{\alpha_{n-2} R_n}, \label{multi-lit-rich}
\end{equation}
in terms of which a multiple tensor product takes the form
\begin{equation}
R_1\otimes \ldots \otimes R_n = \sum_P c^{P}_{R_1\ldots R_n} P, \\
\end{equation}

Finally we define \emph{skew Schur functions}
\begin{equation}
s_{Q/R}=\sum_P c^{Q}_{RP}s_P.
\end{equation}
For trivial representation $R=\bullet$, we have
$$
s_{Q/\bullet}=s_Q.
$$
and
$$
\textrm{If not}\ Q \subset  R \ \  \Leftrightarrow \ \  s_{R/Q}=0. \label{skew-zero}
$$

For Schur functions, we have the following identities
\bea
s_R(c x) &=& c^{|R|} s_{R} (x) \label{schur-c} \nn
s_R(q^{\rho}) &=& q^{\kappa_R /2} s_{R^t} (q^{\rho}) \label{schur-q} \nn
s_R(q^{\rho}) &=& (-1)^{|R|} s_{R^t} (q^{-\rho}) \label{schur-1} \nn
s_Q(q^{\rho})s_R(q^{Q+\rho}) &=& s_R(q^{\rho})s_Q(q^{R+\rho}) \label{schur-change}.
\eea
Skew Schur functions satisfy
\bea
s_{Q/R}(c x) &=& c^{|Q|-|R|} s_{Q/R} (x) \label{skew-c} \nn
s_{Q/R}(q^{\rho}) &=& (-1)^{|Q|-|R|} s_{Q^t/R^t} (q^{-\rho}) \label{skew-1}.
\eea
In addition, we have the summation formulas for Schur functions 
\bea
\sum_{R}  s_{R}(x) s_{R}(y) &=& \prod_{i,j} \frac{1}{1-x_i y_j}  \label{schur-sum3} \nn
\sum_{R}  s_{R}(x) s_{R^t}(y) &=& \prod_{i,j} (1+x_i y_j)  \label{schur-sum4}
\eea
and for skew Schur functions
\bea
\sum_{\eta}  s_{Q/ \eta}(x) s_{R/ \eta}(y) &=& \prod_{i,j} (1-x_i y_j) \ \sum_{\eta} 
s_{\eta/R}(x) s_{\eta/Q}(y) \label{schur-invert1} \nn
\sum_{\eta}  s_{Q/ \eta}(x) s_{R/ \eta}(y) &=&  \prod_{i,j} \frac{1}{1+x_i y_j} \
\sum_{\eta}  s_{\eta^{t}/R}(x) s_{\eta/Q}(y) \label{schur-invert2} \nn
\sum_{\eta}  s_{\eta/R}(x) s_{\eta}(y) &=& s_R(y) \sum_{\mu} s_{\mu}(x)  s_{\mu}(y) 
\label{schur-sum2} \nn
\sum_{\eta}  s_{\eta^t/R}(x) s_{\eta}(y) &=& s_R(y) \sum_{\mu} s_{\mu}(x)  s_{\mu^t}(y) 
\label{schur-sum6} \nn
\sum_{\eta}  s_{R/ \eta}(x) s_{\eta/Q}(y) &=& s_{R/Q}(x,y),  \label{schur-sum5} \nn
\sum_{\eta}  s_{R/ \eta}(x) s_{\eta}(y) &=& s_{R}(x,y)  \label{schur-sum1}.
\eea
the last two sums being  over partitions $\eta$ such that $Q\subset \eta \subset R$.

For the special case a partition with a single row $R=(R_1,0,0,\ldots)$,  the Schur function is
related to the quantum dilogarithm as
\be
s_{R=(R_1,0,\ldots)}(q^{\rho})  =  (-1)^{R_1}q^{R_1^2/2}\xi(q)L\Big((R_1+\frac{1}{2})g_s,q\Big)
\ee 
where 
$$ \xi(q) = \prod_{i=1}^{\infty} {1 \over 1 - q^i} .$$


\section*{Appendix B} \label{induct}

Here we prove the many-row Hopf-link expansion formula (\ref{claim-rhs}) for the simplified case of
$n$ branes on the positive slice only, whose positions determine the values of $a_1,\ldots, a_n$.
 Let us recall, that the normalized crystal partition function in the present case is
\begin{eqnarray}
\tz(a_1,\ldots,a_n) &=&  L(a_1,q)\ldots L(a_n,q) \cdot \label{crys-branes} \\
& & \cdot
\big(1-\frac{a_1}{a_2}\big)\big(1-\frac{a_1}{a_3}\big)\ldots\big(1-\frac{a_1}{a_n}\big)\ldots\big(1-\frac{a_{n-1}}{a_n}\big).
\nonumber
\end{eqnarray}

In this case the statement (\ref{claim-rhs}) takes the form
\begin{equation}
\tz(a_1,\ldots,a_n)=\sum_{R_1,\ldots,R_n} a_1^{R_1}\ldots a_n^{R_n} \
s_{(R_n,R_{n-1},\ldots,R_1)}(q^{\rho}). \label{claim-one-leg}
\end{equation}

We should note, that expansion contains Schur functions corresponding to 'improper'
partitions (with negative number of boxes, or not decreasing in length). But these
are taken into account in the proof below automatically, due to structure of
Schur functions.

We prove (\ref{claim-one-leg}) by induction on number of branes $n$. The first step in
the induction is the expression for the dilogarithm
\begin{equation}
L(a,q)=\sum_{R=0}^{\infty} a^R h_R(q^{\rho}), \label{L-h}
\end{equation}
as a single variable $a$ the sum is over one-row partitions of length $R$,
and $s_{(R,0,\ldots)}=h_R$.

In the second induction step, let us assume that
$\tz (a_1,\ldots,a_n)$ is given by (\ref{claim-one-leg}), and we add one more
brane at $a_0$. Then
\begin{eqnarray}
\tz(a_1,\ldots,a_n,a_{0}) &=& \tz(a_1,\ldots,a_n)L(a_{0},q)
\cdot\nonumber \\
& & \cdot \big(1-\frac{a_1}{a_{0}}\big)\ldots \big(1-\frac{a_n}{a_{0}}\big).
\label{C3-leg-recursive}
\end{eqnarray}

If we expand w.r.t. all $a_i$ and use (\ref{claim-one-leg}) and (\ref{L-h}), the
coefficient at $a_0^{R_0}\cdots a_{n}^{R_{n}}$ is equal to (for now we skip arguments)
$(q^{\rho})$)
\begin{equation}
h_{R_{0}}s_{(R_m,\ldots,R_1)} - h_{R_{0}+1}\sum_{i=1}^{m}s_{(\hat{i})} +
h_{R_{0}+2}\sum_{i\neq j}^{m}s_{(\hat{i},\hat{j})} - \ldots h_{R_{0}+n}s_{(R_n
-1,\ldots, R_1 -1)}, \label{proof-sum-1}
\end{equation}
where $\hat{i}$ means, that $i$'th variable $R_i$ is replaced by $(R_i -1)$, for
example
\begin{equation}
s_{(\hat{i},\hat{j})} = s_{(R_n,R_{n-1},\ldots,R_{i}-1,\ldots,R_{j}-1,\ldots,R_1)}.
\end{equation}
In the first term in this expression no variable is reduced by 1, and in the last
term all $m$ variables are reduced. In other terms several variables are reduced,
and the sums are over all possible combinations of choosing this number of variables
from the set $(R_1,\ldots,R_n)$.

The final observation is that (\ref{proof-sum-1}) is Laplace expansion of the
determinant defining $s_{(R_0,R_n,\ldots,R_1)}$ along the first row (\ref{s-det})
\begin{equation}
s_{(R_0,R_n,\ldots,R_1)} =
\left\| \begin{array}{cccc}
h_{R_0} & h_{R_{0}+1} & \ldots & h_{R_{0}+n} \\
h_{R_{n}-1} & h_{R_{n}} & \ldots & h_{R_{n}+n-1} \\
\vdots & \vdots & \ddots & \vdots \\
h_{R_{1}-m} & h_{R_{1}-n+1} & \ldots  & h_{R_{1}}
\end{array} \right\|  \label{sKm-K1}
\end{equation}
where double lines denote determinant. This completes the induction and proves
(\ref{claim-one-leg}).  In the more general case for branes in both legs a completely analogous
 proof can be constructed.
 \bigskip
 
{\bf \large \noindent Improper partitions}
\smallskip

In the above expansion, it should be stressed that in general not all $W_{PR}$ correspond
 to Hopf link invariants. They correspond only
in the case when $P$ and $R$ are proper partitions, i.e. if $R_n \geq R_{n-1} \geq\ldots
\geq R_1 \geq 0$, and similarly for the representation $P$. Otherwise $W_{PR}$ are just
coefficients resulting from the expansion, but generally these cannot be thought of as Hopf
link invariants. Nonetheless, functions $s_P$ involved in $W_{PR}$ are still given
by the determinant $(\ref{s-det})$, and thus we will call them \emph{improper Schur
functions}.

Moreover, the summations over $R_i$ and $P_i$ in (\ref{claim-rhs})
 don't start from 0, because in the crystal partition function expansions there 
are also terms with negative powers of $a_i,\ b_i$. These negative powers arise only from
prefactors for strings stretched between branes on the same leg, which are of the
form $(1-a_i/a_j)$, and there is always finite number of such terms.

In fact, the easiest way to take care of them is to understand the summations in
(\ref{claim-rhs}) as running over all integers, positive and
negative. The very structure of Schur's functions, together with the fact that
$h_i=0$ for $i<0$, will assure that only relevant terms will be non-zero, and we get
the correct result. In particular, this means that there will be partitions $R$ with
'negative number of boxes' in some rows, $R_i<0$.  So if we expand determinant
(\ref{s-det}) for the corresponding \emph{improper Schur functions} $s_R$, and use
$h_{i<0}=0$, we are left with are Schur functions for partitions with
lower number of rows, now only of positive length. These new functions can also be
proper or not, according to whether lengths of their rows are properly decreasing.

Thus, if we put $n$ branes on one leg, the crystal expansion in fact contains
information about all proper knot invariants, and finite number of improper knot invariants 
for partitions with all number of rows $1,\ldots,n$.


\section*{Appendix C} \label{app-vertex}

In this appendix we introduce A-model topological vertex calculational framework. The most convenient form of the vertex is representation basis, in which vertex amplitudes can be expressed in terms of Schur functions. The general formula for topological vertex in the canonical framing is \cite{ok-re-va}
\begin{equation}
C_{R_1 R_2 R_3} = q^{\frac{1}{2}(\kappa_{R_2}+\kappa_{R_3})} s_{R_{2}^{t}}(q^{\rho})
\, \sum_{P} s_{R_{1}/P}(q^{ R_{2}^{t}+\rho}) s_{R_{3}^{t}/ P}(q^{ R_{2}+\rho}).
\label{vertex}
\end{equation}

The crucial property of $C_{R_1 R_2 R_3}$ in the canonical framing is cyclicity
w.r.t. representations $R_i$. The above formula also immediately implies
\begin{equation}
C_{R_1 R_2 R_3} = q^{\frac{1}{2}\sum_{i} \kappa_{R_i}} C_{R_{1}^{t} R_{3}^{t}
R_{2}^{t}}. \label{C-transpose}
\end{equation}

The identities from appendix \ref{app-schur} lead to the following special cases, with some representations involved being trivial $\bullet$
\begin{eqnarray}
C_{R \bullet \bullet} & = & q^{\kappa_{R}/2} s_{R^{t}}(q^{\rho}) = s_{R}(q^{\rho}), \label{C-0R0} \\
C_{PR\bullet} & = & q^{\frac{1}{2}\kappa_R}s_{P}(q^{\rho})s_{R^t}(q^{\rho+P}) = \label{C-PR} \\
& = & q^{\frac{\kappa_P}{2}} \sum_{\eta} s_{R/ \eta}(q^{\rho}) s_{P^{t}/
\eta}(q^{\rho}). \label{C-PR-skew}
\end{eqnarray}

The vertex with one trivial representation is closely related to the leading term of the Hopf Link invariant $W_{PR}$, which also can be expressed in terms of Schur functions
\begin{eqnarray}
W_{PR}&=& q^{\kappa_{R}/2}C_{PR^t\bullet} = \label{W-C} \\
& = & s_{P}(q^{\rho})s_R(q^{\rho+P})= \label{W-schur} \\
& = & q^{\frac{1}{2}(\kappa_{P}+\kappa_{R})} \sum_{\eta} s_{R^t/ \eta}(q^{\rho})
s_{P^{t}/ \eta}(q^{\rho}),
\end{eqnarray}
and it is not difficult to show that
\be
W_{PR}(q) = (-1)^{|P|+|R|} W_{P^t R^t}(q^{-1}). \label{W-PtRt}
\ee

The important feature of the vertex is a framing ambiguity, which arises as a need to specify an
integer number for each stack of branes on a leg of $\mathbb{C}^3$. The vertex in a particular framing specified by numbers $f_1, f_2, f_3$ corresponding to representations $R_i$ on different axes is given as
\begin{equation}
C^{f_1,f_2,f_3}_{R_1 R_2 R_3} = (-1)^{\sum_i f_i |R_i|} q^{\sum_i f_i \kappa_{R_i}
/2} C_{R_1 R_2 R_3}, \label{framing}
\end{equation}
where $|R_i|$ denotes number of boxes in the Young diagram for a given representation. The canonical framing (\ref{vertex}) corresponds to $f_i=0$.

It is also possible to reverse orientation of the branes on one leg, what can 
be interpreted as changing branes to antibranes. To obtain vertex amplitude 
with an antibrane on the first axis one should substitute
\begin{equation}
C_{PQR}\ \to \ (-1)^{|P|} C_{P^t QR}, \label{vertex-orient}
\end{equation}
and similarly for any other leg.

To construct the full toric diagram, one has to glue together $\mathbb{C}^3$ 
patches. Gluing together just two patches gives a resolved conifold with Kahler parameter $Q=q^N=e^{-t}$, and the propagator is given by $(-Q)^{|R|}$. The orientations of two glued axes must be consistent, and sum over representations performed, what leads to
\begin{eqnarray}
Z^{P^1} & = & \sum_R C_{\bullet \bullet R^t} (-Q)^{|R|} C_{R \bullet\bullet}  = \sum_R s_{R}(q^{-\rho}) s_{R}(Qq^{\rho}) = \nonumber \\
& = & \prod_{i,j=1}^{\infty} \frac{1}{1-Qq^{i-j}} =
\exp\Big(-\sum_{k=1}^{\infty}\frac{Q^{k}}{k[k]^2} \Big). \label{Z-P1}
\end{eqnarray}

\bigskip
It is also possible, though a bit more complicated, to consider branes on internal legs of toric diagram and configurations of several stacks of branes on one leg. In the case of one stack of branes on a compact leg one more parameter $d=g_s D$ should be introduced, which denotes the position of the brane along the compact leg, as measured from the left vertex. To properly glue two vertexes with additional brane between them, two additional summations must be introduced representing strings ending on the brane from the left and from the right, so the relevant vertex factor takes the form
\begin{equation}
\sum_{R,Q^L,Q^R} C_{\bullet\bullet R\otimes Q^L} (-1)^s q^f e^{-L} C_{R^t\otimes
Q^R\bullet\bullet}\, Tr_{Q^L }V\, Tr_{Q^R}V^{-1}  \label{comp-leg-brane}
\end{equation}
where a framing of the brane $p$ has also been taken into account, so that
\bea
L & = & |R|t + |Q^L|d + |Q^R|(t-d), \label{distances-L} \\
f & = & \frac{p}{2} \kappa_{R\otimes Q^L} + \frac{n+p}{2} \kappa_{R^t\otimes Q^R}, \nn
s & = & |R| + p|R\otimes Q^L| + (n+p)|R^t \otimes Q^L|.
\eea
The additional number $n=|v' \times v|$ is determined by planar directions of two axes $v$ and $v'$ of glued vertexes, and in all cases we consider it equals zero.

For more stacks of branes we need to specify a position of each stack as $d_i=g_s D_i$.  The holonomy
matrix corresponding to branes at $d_i$ is denoted $V_i$. In fact, to get agreement with crystal results we will need to absorb $d_i$ into $V_i$.  Moreover, in this case we have to choose different representations $R_i$ for each stack of branes, and for a given leg of the vertex consider the tensor product of representations $C_{P,Q,\otimes_i R_i}$. In addition, for each pair of branes at $d_i,d_j$ we have to introduce an additional factor from strings stretched between them
\begin{equation}
\sum_P (-1)^{|P|}\, Tr_P V_i\, Tr_{P^t} V_j^{-1}. \label{stretched}
\end{equation}

If there are several branes on an internal leg, also summations from the left and right vertexes for each brane must be introduced, as well as an overall summation over $|R|$ as in (\ref{comp-leg-brane}).

It is also important to make clear how summations over tensor products should 
be understood. The Hopf Link with a single factor of $|P_1 \otimes P_2|$ can be obtained from a fusion rule
\begin{equation}
W_{P_1\otimes P_2, R} = \sum_{\alpha} c^{\alpha}_{P_1 P_2} W_{\alpha R}  =  \sum_{\alpha} q^{\frac{\kappa_{\alpha}+\kappa_R}{2}} c^{\alpha}_{P_1 P_2} s_{\alpha^t/\eta}(q^{\rho}) s_{R^t/\eta}(q^{\rho}), \label{W-tensor}
\end{equation}
and then related to topological vertex by (\ref{W-C}). When a few factors of $|P_1 \otimes P_2|$ appear, the internal summation over $\alpha$ should also be introduced\footnote{We thank Marcos Marino 
for explaining this point.}, with each such factor replaced by $\alpha$. For example, for two stacks of branes on one leg of $\mathbb{C}^3$ in $(-1)$ framing we have
\begin{displaymath}
C_{P_1\otimes P_2, R, \bullet} (-1)^{-|P_1\otimes P_2|} q^{-\frac{\kappa_{P_1\otimes P_2}}{2}} = \sum_{\alpha}  c^{\alpha}_{P_1 P_2} s_{\alpha^t/\eta}(q^{\rho}) s_{R/\eta}(q^{\rho}) (-1)^{|\alpha|} =
\end{displaymath}
\begin{equation}
= (-1)^{|P_1|+|P_2|} \sum_{\alpha} c^{\alpha}_{P_1 P_2} s_{\alpha^t/\eta}(q^{\rho}) s_{R/\eta}(q^{\rho}), \label{comp-brane-tensor-1}
\end{equation}
where two factors of $q^{\kappa_{\alpha} /2}$ (from vertex expression and $(-1)$ framing) canceled each other, and formulas (\ref{lit-rich-prop}) and (\ref{lit-rich-nonzero}) have been used.

The internal summation arising in quantities with tensor product involved is a crucial and subtle issue. In particular, knot invariants in different framings but without tensor product differ just by an overall sign and factors of $q$. On 
the other hand, the summation implicit in tensor product formulae changes the structure of polynomials representing knot invariants. For example, in canonical framing we have
\begin{equation}
C_{1\otimes 1, 1, \bullet} = W_{1\otimes 1, 1, \bullet} = \frac{(q^2-q+1)^2}{(q-1)^3\sqrt{q}}.
\end{equation}
On the other hand, in framing $(-1,0)$
$$
C_{1\otimes 1, 1, \bullet} (-1)^{|1\otimes 1|}q^{-\frac{\kappa_{1\otimes 1}}{2}}=  W_{1\otimes 1, 1, \bullet} (-1)^{|1\otimes 1|}q^{-\frac{\kappa_{1\otimes 1}}{2}} =
$$
\begin{equation}
=  W_{1,2}\, q^{-1} + W_{1,2^t}\, q = \frac{(2q^2-3q+2)\sqrt{q}}{(q-1)^3},
\end{equation}
and this is also precisely the coefficient which we get from crystal expansion without the factor (\ref{stretched}) $(1-\frac{a_1}{a_2}) $ (and up to $q$ inversion) 
with two branes at $a_1, a_2$ on one slice of the crystal and antibrane on the other slice.

\end{document}